\documentclass[floatfix,showpacs,prl,aps,twocolumn]{revtex4}
\usepackage{bm}
\usepackage{amsmath}
\usepackage{amssymb}
\usepackage{latexsym} 
\usepackage{amsfonts} 
\usepackage{epsfig}
\usepackage{psfrag}

\newcommand{\ket}[1]{|#1\rangle}

\newcommand{\nn}{\nonumber\\} 
\newcommand{\bea}{\begin{eqnarray}}
\newcommand{\ea}{\end{eqnarray}}

\begin{document}


\title{Quantum simulator for the Ising model with electrons floating
  on a helium film} 


\author{Sarah Mostame$^1$ and Ralf Sch\"utzhold$^{1,2,*}$}

\affiliation{$^1$Institut f\"ur Theoretische Physik, 
Technische Universit\"at Dresden, 01062 Dresden, Germany
\\
$^2$Fachbereich Physik, Universit\"at Duisburg-Essen, 
D-47048 Duisburg, Germany}

\begin{abstract} 
We propose a physical setup that can be used to simulate the quantum
dynamics of the Ising model in a transverse field 
with present-day technology. 
Our scheme consists of electrons floating on superfluid helium which
interact via Coulomb forces. 
In the limit of low temperatures, the system will stay near its ground 
state where its Hamiltonian is equivalent to the Ising model and thus 
shows phenomena such as quantum criticality. 
Furthermore, the proposed design could be generalized in order to
study interacting field theories (e.g., $\lambda\phi^4$) and adiabatic 
quantum computers. 
\end{abstract} 


\maketitle
 
 
{\em Introduction}\quad
Richard Feynman's observation \cite{feynman} that classical computers 
cannot effectively simulate quantum systems bred widespread interest 
in quantum computation.
He thought up the idea of a quantum processor which uses the effects of
quantum theory instead of classical physics. 
As an example, Feynman proposed a {\em universal quantum simulator} 
consisting of a lattice of spins
with nearest neighbor interactions that are freely specifiable and can
efficiently reproduce the dynamics of {\em any} other many-particle
quantum system with a finite-dimensional state space \cite{feynman}. 
Although such universal quantum computers of sufficient size 
(e.g., number of QuBits, i.e., spins) are not available yet, it is
possible to design a special quantum system in the laboratory which
simulates the quantum dynamics of a particular model of interest. 
Such a designed quantum system can then be regarded as a special
quantum computer (instead of a universal one, which is more challenging) 
which just performs the desired quantum simulation, see, e.g., 
\cite{porras,ralf,byrnes}.

In the following, we present a design for a quantum simulator for the
Ising spin chain in a transverse field and demonstrate that it could
be feasible with present-day technology, i.e., electrons floating on a
thin superfluid Helium film. 
A similar idea based on trapped ions has been pursued in \cite{porras}. 
Nevertheless, since different experimental realizations possess distinct 
advantages and drawbacks, it is still  worthwhile to study an alternative 
set-up.
For example, the number of coherently controlled ions in a trap is
rather limited at present, whereas our proposal can be scaled up to a
large number of electrons more easily -- which is important for 
exploring the continuum limit and scaling properties etc. 

{{\em The model}}\quad
We want to simulate the quantum dynamics of the one-dimensional Ising
chain consisting of $n$ spins with nearest-neighbor interaction $J$
plus a transverse field $\Gamma$ along the $x$-direction ($\hbar=1$) 
\bea
\label{Hamiltonian}
{H} 
= 
-\sum_{j=1}^n\left\{\Gamma\,\sigma^x_j+
\,J \,\sigma^z_j\sigma^z_{j+1}\right\} 
\,, 
\ea
where \mbox{${\bm\sigma}_j=(\sigma^x_j,\sigma^y_j,\sigma^z_j)$} are
the spin-1/2 Pauli matrices acting on the $j$th qubit. 
This model has been employed in the study of quantum phase transitions
and percolation theory \cite{sachdev}, spin glasses
\cite{sachdev,fischer}, as well as quantum annealing 
\cite{chakrabarti,santoro-kadowaki} etc. 
Although the Hamiltonian (\ref{Hamiltonian}) is quite simple and can
be diagonalized analytically, the Ising model is considered a
paradigmatic example \cite{sachdev} for second-order quantum phase
transitions and is rich enough to display most of the basic phenomena
near quantum critical points. 
For $\Gamma \gg J$, the ground state is paramagnetic 
\mbox{$\ket{\rightarrow\rightarrow\rightarrow\dots}$} 
with all spins polarized along the $x$ axis. 
In the opposite limit $\Gamma \ll J$, the nature of the ground state(s) 
changes qualitatively and there are two degenerate ferromagnetic phases 
with all spins pointing either up or down along the $z$ axis 
$\ket{\uparrow\uparrow\uparrow\dots}$ or 
$\ket{\downarrow\downarrow\downarrow\dots}$. 
The two regimes are separated by a quantum phase transition 
at the critical point $\Gamma_{\rm cr}=J$, where the excitation gap
vanishes (in the thermodynamic limit $n\uparrow\infty$) and the
response time diverges. 
As a result, driving the system through its quantum critical point at
a finite sweep rate entails interesting non-equilibrium phenomena such 
as the creation of topological defects, i.e., kinks \cite{dziarmaga}. 
Furthermore, the transverse Ising model can also be used to study the
order-disorder transitions at zero temperature driven by quantum
fluctuations \cite{chakrabarti,sachdev}.
Finally, two-dimensional generalizations of the Ising model can be
mapped onto certain adiabatic quantum algorithms 
(see, e.g., \cite{dwave}). 
However, due to the evanescent excitation energies, such a phase 
transition is rather vulnerable to decoherence, which must be taken 
into account \cite{deco}.

\begin{figure}
\includegraphics[height=5cm]{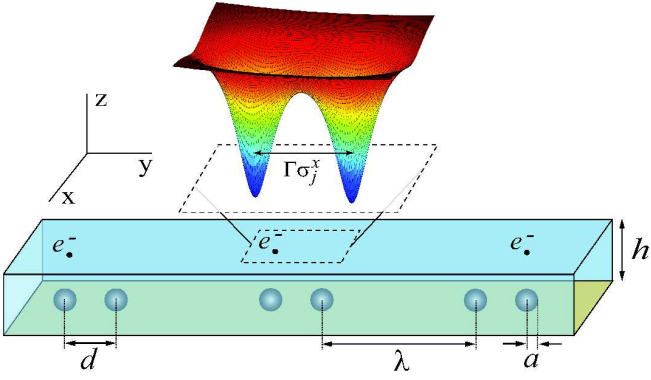} 
\caption{\label{Energy}
{Sketch of the proposed analogue quantum simulator.
Electrons ($e^-$) are floating on a low-temperature helium film  
of height $h$ adsorbed on a silicon substrate.
A double-well potential for each single electron is created by a pair
of golden spheres of radius $a$ and distance $d$ on the bottom of the
helium film. 
The double wells at each site provide two lowest states of the
electron and model the spin states $\ket{\uparrow}$ and
$\ket{\downarrow}$ at each site $j$. 
The tunneling rate between the two wells corresponds to 
the transverse field term \mbox{$\Gamma \sigma_j^x$}. 
The electrons are lined up at distances $\lambda$ and interact via
Coulomb forces, which creates the term $J\sigma^z_j\sigma^z_{j+1}$. 
}} 
\label{Analogue}
\end{figure} 

{{\em The analogue}}\quad
In order to reproduce the quantum dynamics of the 1+1 dimensional
Ising model (\ref{Hamiltonian}), we propose trapping a large number of
electrons on a low-temperature helium film of thickness $h$ 
(e.g., $h=110$ nm) adsorbed on a silicon substrate \cite{leiderer}.
Due to the polarizability $\varepsilon\approx1.06$ of the Helium film,
the electrons are bound to its surface (i.e., in $z$-direction) via
their image charges and the large potential barrier (around 1 eV) for
penetration into the helium film \cite{andrei}. 
Since the binding energy of around 8 K is much larger than the
temperature $T$ (below 1 K) and the width of the electron wave packet
in $z$-direction (of order 8 nm) is much smaller than all other
relevant length scales, the electron motion is approximately
two-dimensional ($x,y$-plane). 

\begin{figure}
\includegraphics[height=4cm]{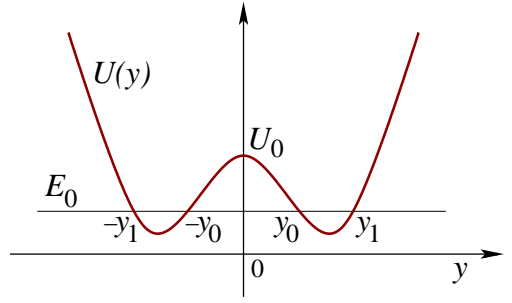} 
\caption{\label{Turning-Points}
Sketch of the double-well potential $U(y)$ with four turning points
  for the energy $E_0$.} 
\end{figure}

In our scheme, each single electron on top of the helium film is 
trapped by a pair of golden spheres of radius $a$ (e.g., $a=10$ nm)
and distance $d$ (e.g., $d=60$ nm) attached to the silicon substrate 
(i.e., on the bottom of the helium film, cf.~Fig.~\ref{Analogue}). 
Depending on its position $x,y$, the electron will also induce image
charges in the two golden spheres (which act as a pair of quantum
dots) and hence experience a double-well potential
\bea
\label{Potential}
U_{\rm{w}}(x,y)
=
-\,
\frac{ae^2
\left(
x^2+y^2+\alpha^2+\beta^2
\right)/4\pi\varepsilon}
{
\left(
x^2+y^2+\alpha^2+\beta^2
\right)^2 
-4\alpha^2 y^2}
\,
,
\ea
with $\alpha=d/2+a$ and $\beta^2=h^2-a^2$.
Since this potential is quite deep and symmetric $U(x,y)=U(x,-y)$,
cf.~Fig.~\ref{Turning-Points}, the ground state
wave-function $\psi_{\rm{S}} (x,y)$ is given by the symmetric
superposition of the two Wannier states $\psi_0(x,\pm y)$ while the
first excited state $\psi_{\rm{A}} (x,y)$ is the anti-symmetric
combination
\bea
\label{Symmetric-Antisymmetric}
\psi_{\rm{S}} (x,y)
=
\frac{\psi_0(x,y)+\psi_0(x,-y)}{\sqrt{2}}
\, 
\to
\,
\frac{\ket{\uparrow}+\ket{\downarrow}}{\sqrt{2}}
\,,
\nn
\psi_{\rm{A}} (x,y)
=
\frac{\psi_0(x,y)\,-\,\psi_0(x,-y)}{\sqrt{2}}
\, 
\to
\,
\frac{\ket{\uparrow}-\ket{\downarrow}}{\sqrt{2}}
\, .
\ea
For a sufficiently high potential barrier between the two wells, the 
Wannier state $\psi_0(x,y)$ is strongly concentrated in the left well
and models the spin state $\ket{\uparrow}$ and vice versa. 
The tunneling between the two states is then described by the Pauli
operator $\sigma^x$ with $\sigma^x\ket{\uparrow}=\ket{\downarrow}$ and 
$\sigma^x\ket{\downarrow}=\ket{\uparrow}$ such that the tunneling rate, 
given by the difference of the eigenenergies $E_A-E_S$ of $\psi_{\rm{S}}$
and $\psi_{\rm{A}}$, corresponds to the transverse field $\Gamma$ in
Eq.~(\ref{Hamiltonian}). 
In the limit of strong localization (i.e., weak tunneling), the energy
splitting $E_A-E_S$ between the two levels can be estimated via the WKB
approximation \cite{razavy}
\bea
\label{WKB}
E_{\rm{A}}-E_{\rm{S}}\approx\frac{\omega}{\pi}
\exp\left[-\int_{-y_0}^{y_0} dy\,\left|p(y)\right|\right]
\,.
\ea
Here $\omega$ is the oscillation frequency (within one well) and 
$\pm y_0$ are the two inner (classical) turning points, 
cf.~Fig.~\ref{Turning-Points}. 
The integrand is given by 
$p(x,y)=\sqrt{2m_e\left[E_0-U(x,y)\right]}$, where we can set $x=0$
since the tunneling probability away from the $x=0$-axis is strongly
suppressed. 
Finally, the energy $E_0$ determines the turning points and $m_e$ is
the electron mass. 
For the parameters above, each valley can well be approximated by a
harmonic oscillator 
\bea
\label{Oscillator}
U_{\rm{w}} (x,y\approx\pm y_{\rm min}) 
\approx 
\frac{ae^2}{4\pi\varepsilon\beta^4} 
(x^2+[y\mp y_{\rm min}]^2)
\,,
\ea
and thus we obtain 
$E_0 \approx \sqrt{ae^2/2\pi\varepsilon m_e\beta^4} \approx \omega$. 

So far, we derived the term $\Gamma\sigma_j^x$ in
Eq.~(\ref{Hamiltonian}) via Eqs.~(\ref{Potential}),
(\ref{Symmetric-Antisymmetric}), (\ref{WKB}), and (\ref{Oscillator}). 
In order to simulate the remaining part, we propose to line up the pairs
of quantum dots at equal distances $\lambda$ (e.g., $\lambda=600$ nm),
where the parameters are supposed to obey the following hierarchy
\bea
\label{Hierarchy}
\lambda 
\gg
h
>
d
\gg
a
\,.
\ea
In this limit, the interaction between the electrons will be dominated
by the direct Coulomb repulsion between nearest neighbors 
$U_{\rm{c}}(x,y)=\sum_{j=1}^n U_{\rm{c}}^{j,j+1}$ with $n$ 
denoting the number of electrons floating on the helium film. 
For $\lambda \gg d$, we may Taylor expand the Coulomb interaction into
powers of $y/\lambda$ due to $y\approx\pm d/2$.  
The zeroth-order term is constant and thus irrelevant while the
first-order contributions vanish (up to boundary terms) after the sum
over sites $j$. 
Thus, the leading term is bilinear in the electron positions 
\bea
\label{Coulomb}
U_{\rm{c}}(x,y)
\approx
-\frac{e^2}{2\pi\varepsilon_0\left(\lambda +d+4a\right)^3
}\sum_{j=1} ^n y_j y_{j+1}
\,,
\ea
and precisely corresponds to the $J\sigma^z_j\sigma^z_{j+1}$ term in 
Eq.~(\ref{Hamiltonian}) with the effective coupling 
\bea
J=\frac{e^2(d+2a)^2}{8\pi\varepsilon_0(\lambda + d+ 4a)^3}
\,.
\ea
%

{\em Experimental parameters}\quad
For the example values given in the text, we obtain 
$\Gamma\approx$~0.1~K for the tunneling rate and the same value 
$J\approx$~0.1~K for the effective coupling, i.e., we are
precisely in the quantum critical regime.
However, deviations from this critical point should be easy to realize
experimentally by varying the height $h$ of the helium film, since the
tunneling rate depends strongly (in fact, exponentially) on $h$,
whereas the Coulomb force remains approximately constant. 
In order to see quantum critical behavior, i.e., to avoid thermal
fluctuations, the temperature should ideally be well below this value
0.1~K (or at least not far above it). 

Furthermore, the Coulomb repulsion energy between two electrons
(zeroth-order term) of about 11~K would tend to destabilize the
electron chain.
Fortunately, this effect is compensated by the binding energy between
the electron and its image on the sphere, which is around 13~K and
thus stabilizes the electron chain. 
The probability for the electron to penetrate the helium film by
tunneling to one of the golden spheres is extremely small 
(of order $10^{-16}$) and can be neglected. 
Finally, the ground-state energy $E_0\approx$~1.4~K
(within the harmonic oscillator approximation) 
is reasonably well below the barrier height $U_0\approx$~3.1~K 
such that the WKB approximation should provide a reasonable estimate.  
(The tunneling probability of 0.08 is also small enough.) 
On the other hand, $E_0\approx$~1.4~K is a measure of the distance
between the two lowest-lying states in
Eq.~(\ref{Symmetric-Antisymmetric}) and the remaining excited states
in the double-well potential.
As a result, these additional states do not play a role for
temperatures well below one Kelvin and thus the Hamiltonian
(\ref{Hamiltonian}) provides the correct low-temperature
description. 

{{\em Read-out scheme}}\quad
Having successfully simulated the Ising Hamiltonian (\ref{Hamiltonian}),
one is lead to the question of how to actually measure its properties, 
e.g., how to detect signatures of quantum critical behavior. 
As one possibility, let us imagine enclosing the Ising chain symmetrically 
by two electrodes in the form of spheres of radius $R=100~\mu\rm m$ and a 
distance of 1~mm aligned along the chain axis.  
Applying a voltage of $1\mu$V, an approximately constant electric field 
of $4\times10^{-4}$~V/m acts on all the electrons and induces the 
perturbation Hamiltonian 
\bea
\label{perturbation-Hamiltonian}
{H}_{\rm pert}
= 
\sum_{j=1}^n\gamma\,\sigma^z_j
\,, 
\ea
corresponding to a longitudinal field (in addition to the 
transversal one $\Gamma\,\sigma^x_j$). 
For $d=60~\rm nm$, we get $\gamma\approx0.1~\mu\rm K$, 
i.e., a very weak perturbation $\gamma\ll\Gamma$.

Deep in the paramagnetic phase $\Gamma\gg J$, the response of the 
system to this weak perturbation $\gamma\ll\Gamma$ is rather small 
$\langle\sigma^z_j\rangle\approx\gamma/\Gamma$.
Approaching the phase transition, however, the static susceptibility 
$\chi_\gamma=\lim_{\gamma\to0}\langle\sigma^z_j\rangle/\gamma$ 
grows and finally diverges at the critical point. 
In the broken symmetry phase, the perturbation 
(\ref{perturbation-Hamiltonian}) lifts the degeneracy 
$\sigma^z_j\to-\sigma^z_j$ and hence the response is non-analytic,
i.e., independent of the smallness of $\gamma$: 
e.g., for $J\gg\Gamma$, we have 
$\langle\sigma^z_j\rangle={\rm sign}(\gamma)=\pm1$.
This signal $\langle\sigma^z_j\rangle$ indicating the phase 
transition can be picked up by the two electrodes for which 
the Ising chain acts like a dielectric medium and induces a 
voltage shift of order nano-Volt per electron (for $J\gg\Gamma$), 
which should be measurable for a sufficiently large number of sites. 
In addition to the static case, one could also study the 
time-resolved response $\langle\sigma^z_j(t)\rangle$ to a 
varying voltage $\gamma(t')$, which is determined by the 
dynamical correlator $\langle\sigma^z_i(t')\sigma^z_j(t)\rangle$ 
in lowest-order response theory.

Even in the absence of an externally imposed voltage, the chain 
induces spontaneous voltage fluctuations in the electrodes, 
which are strongest (of order nano-Volt per electron) deep in the 
ferromagnetic phase. 
The variance of these fluctuations yields the correlator sum 
$\sum_{ij}\langle\sigma^z_i\sigma^z_j\rangle$ which is 
an order parameter for the phase transition and allows us to 
detect topological defects (i.e., kinks) which might have been 
produced during the sweep to the ferromagnetic phase: 
In the presence of a kink, the ground-state signal 
$\sum_{ij}\langle\sigma^z_i\sigma^z_j\rangle=n^2$ 
is drastically reduced 
(in average to $\sum_{ij}\langle\sigma^z_i\sigma^z_j\rangle=n^2/3$)
depending on the kink position.
If the kink is precisely in the middle of the Ising chain, 
we get a vanishing signal 
$\sum_{ij}\langle\sigma^z_i\sigma^z_j\rangle=0$, whereas a kink near 
the boundaries does not diminish the signal strongly. 

{{\em Disorder and decoherence}}\quad
In a real experimental set-up, the Hamiltonian will not be exactly 
equivalent to (\ref{Hamiltonian}) due to imperfections such as electric 
stray fields, variations in the film thickness $h$ and further geometric 
parameters $a$, $d$, and $\lambda$ etc. 
Therefore, the original expression (\ref{Hamiltonian}) will typically 
be altered to 
\bea
\label{disorder}
H 
= 
-\sum_{j=1}^n\left\{\Gamma_j\,\sigma^x_j+
\,J_j\,\sigma^z_j\sigma^z_{j+1}+\gamma_j\sigma^z_j\right\} 
\,, 
\ea
where $\Gamma_j=\bar\Gamma+\delta\Gamma_j$ and $J_j=\bar J+\delta J_j$.  
Assuming that the disorder parameters $\delta\Gamma_j$, $\delta J_j$, 
and $\gamma_j$ are much smaller than the excitation gap $\Delta=2|J-\Gamma|$ 
of the undisturbed system (in the continuum limit), the impact of these 
imperfections will be suppressed.
Near the critical point $J\approx\Gamma$, however, this argument fails. 
Still, for a finite number $n$ of electrons, one retains a minimum gap 
(within the symmetric or anti-symmetric subspace, respectively) 
of order $J/n$.
Exploiting this gap might be suitable for a reasonably small systems, but 
for $n\geq100$ electrons, the required accuracy on the sub-percent level is 
probably hard to achieve experimentally. 
E.g., decreasing the diameter of the golden spheres by ten percent with 
the other values remaining the same as before, the tunneling rate 
increases by fifty percent. 

For a sufficiently large number of electrons, the disorder induced by 
imperfections will become relevant near the critical point 
(in one spatial dimension) in view of the critical exponent $\nu=1$ 
of the Ising model, see, e.g., \cite{sachdev}.
(I.e., the renormalization flow is directed away from the homogeneous 
situation.)
In this case, one would expect effects such as local paramagnetic regions 
inside the global ferromagnetic phase and percolation transitions etc. 
Therefore, turning this drawback into an advantage, one might generate 
these imperfections on purpose in order to study the impact of disorder 
onto the phase transition. 
In contrast to the original Hamiltonian (\ref{Hamiltonian}), the above 
form (\ref{disorder}) is no longer analytically solvable and hence much 
less is known about its properties. 
Finally, in a real set-up, the system will also experience decoherence 
due to the inevitable coupling to the environment \cite{deco}.
These effects could be incorporated by operator-valued variations 
$\delta\Gamma_j$, $\delta J_j$, and $\gamma_j$ associated to the 
degrees of freedom of the environment -- where the same arguments apply 
as before. 

{{\em Summary}}\quad
We have proposed a design for the simulation of the quantum Ising
model with a system of electrons floating on a liquid helium film
adsorbed on a silicon substrate.
Since the energy level splitting (tunneling rate $\Gamma$) depends 
exponentially on the thickness of the helium film $h$, 
we may tune the system through the quantum phase transition by
changing $h$ -- which might even be feasible in a time-dependent
manner, cf.~\cite{dziarmaga}. 
The created topological defects (kinks) could be detected via a strong 
reduction of the spontaneous voltage fluctuations in comparison with 
the homogeneous ferromagnetic phase. 

Furthermore, a suitable generalization to two spatial dimensions might
be relevant for adiabatic quantum algorithms, see, e.g.,
\cite{dwave}. 
Note that the realization of a sequential quantum computer based on a 
set of electrons floating on a helium film has been proposed in 
\cite{platzman}. 
In contrast, our proposal is not suited for universal computations, 
but (as one would expect) should be easier to realize experimentally. 

Exploring a different limit, where many eigenstates of the double-well
potential contribute, the proposed set-up could simulate the lattice 
version of interacting field theories such as the $\lambda\phi^4$-model 
in 1+1 dimensions. 

S.~M.~acknowledges fruitful discussions with R.~Farhadifar and
R.~S.~is indebted to G.~Volovik and P.~Leiderer for valuable
conversations.  
This work was supported by the Emmy-Noether Programme of the
German Research Foundation (DFG) under grant SCHU~1557/1-2,3 
and by DFG grant SCHU~1557/2-1.

$^*$ email: {\tt ralf.schuetzhold@uni-due.de}


\begin{thebibliography}{9999}

\bibitem{feynman} 
R.~P.~Feynman
Int.~J.~Theor.~Phys. {\bf 21}, 467 (1986);
R.~P.~Feynman
Found.~Phys. {\bf 16}, 507 (1982).

\bibitem{porras} 
D.~Porras and J.~I.~Cirac,
Phys.~Rev.~Lett.~{\bf 92}, 207901 (2004).

\bibitem{ralf} 
R.~Sch\"utzhold and S.~Mostame,
JETP~Lett.~{\bf 82}, 248 (2005). 

\bibitem{byrnes} 
T.~Byrnes {\em et al}.,
Phys.~Rev.~Lett.~{\bf 99}, 016405 (2007). 

\bibitem{sachdev} 
S.~Sachdev, 
{\em Quantum Phase transitions},
(Cambridge University Press, Cambridge, UK, 1999).

\bibitem{fischer} 
K.~H.~Fischer, and J.~A.~Hertz
{\em Spin glasses},
(Cambridge University Press, Cambridge, UK,  1993).

\bibitem{chakrabarti} 
A.~Das, and B.~K.~Chakrabarti, 
{\em Quantum Annealing and Related Optimisation Methods},  
(LNP~{\bf 679}, Springer-Verlag, Heidelberg 2005).

\bibitem{santoro-kadowaki} 
G.~E.~Santoro {\em et al}.,  
Science~{\bf 295}, 2427 (2002);
T.~Kadowaki, and H.~Nishimori,
Phys.~Rev.~E {\bf 58}, 5355 (1998).

\bibitem{dziarmaga} 
J.~Dziarmaga,
Phys.~Rev.~Lett.~{\bf 95}, 245701 (2005).

\bibitem{dwave} 
IEEE Spectrum online, Tech Talk, February 13th (2007).

\bibitem{deco} 
S.~Mostame, G.~Schaller, and R.~Sch\"utzhold,
Phys.\ Rev.\ A {\bf 76}, R030304 (2007).


\bibitem{leiderer} 
J.~Angrik {\em et al}.,
Journal of Low Temperature Physics, {\bf 137}, 	335 (2004).

\bibitem{andrei} 
E.~Y.~Andrei, Ed.,
{\em Two Dimensional Electron Systems on Helium and Other Cryogenic 
Substrates},
(Academic Press, New York, 1991).

\bibitem{razavy} 
M.~Razavy, 
{\em Quantum theory of tunneling},
(World Scientific, 2003).

\bibitem{platzman} 
P.M.~Platzman and M.I.~Dykman, 
Science {\bf 284}, 1967 (1999). 

\end{thebibliography}
\end{document}